\documentclass[aps,pre,twocolumn,groupedaddress,showpacs,floatfix]{revtex4}
\usepackage{amsmath}
\usepackage{amssymb}
\usepackage{graphicx}

\begin{document}

\title{Exploring Complex Networks through Random Walks}

\author{Luciano da Fontoura Costa} 
\affiliation{Instituto de F\'{\i}sica de S\~ao Carlos. 
Universidade de S\~ ao Paulo, S\~{a}o Carlos, SP, PO Box 369,
13560-970, phone +55 16 3373 9858,FAX +55 16 3371 3616, Brazil,
luciano@if.sc.usp.br}

\date{12th April 2006}

\begin{abstract}   

Most real complex networks -- such as protein interactions, social
contacts, the internet -- are only partially known and available to
us.  While the process of exploring such networks in many cases
resembles a random walk, it becomes a key issue to investigate and
characterize how effectively the nodes and edges of such networks can
be covered by different strategies.  At the same time, it is
critically important to infer how well can topological measurements
such as the average node degree and average clustering coefficient be
estimated during such network explorations.  The present article
addresses these problems by considering random and Barab\'asi-Albert
(BA) network models with varying connectivity explored by three types
of random walks: traditional, preferential to untracked edges, and
preferential to unvisited nodes.  A series of relevant results are
obtained, including the fact that random and BA models with the same
size and average node degree allow similar node and edge coverage
efficiency, the identification of linear scaling with the size of the
network of the random walk step at which a given percentage of the
nodes/edges is covered, and the critical result that the estimation of
the averaged node degree and clustering coefficient by random walks on
BA networks often leads to heavily biased results.  Many are the
theoretical and practical implications of such results.
\end{abstract}

\maketitle

\emph{The crew of the caravel `Nina' also saw signs of land, and a 
small branch covered with berries. Everyone breathed afresh and
rejoiced at these signs.}  

(Christopher Columbus)
\vspace{0.5cm}

\section{Introduction}

Despite its relatively young age, the area of investigation going by
the name of \emph{complex networks}~\cite{Albert_Barab:2002,
Dorog_Mendes:2002, Newman:2003, Boccaletti:2005, Costa_surv:2005} has
established itself as a worthy relative -- or perhaps inheritor -- of
graph theory and statistical physics.  Such a success has been a
direct consequence of the emphasis which has been given to structured
interconnectivity, statistical formulations, interest in applications
and, as in more recent developments (e.g.~\cite{Newman:2003,
Boccaletti:2005}), the all-important paradigm relating structure and
dynamics.  Yet, almost invariably, the analyzed networks are assumed
to be completely known and accessible to us.  Indeed, while so many
important problems involving completely described networks -- such as
community finding (e.g.~\cite{Newman:2004}) -- remain as challenges in
this area, why should one bother to consider incompletely specified
networks?

Perhaps a good way to start making sense of this question is by
considering our future.  To what restaurant are we going tomorrow?
What article will we read next?  Which mirrors will ever see our faces
again?  Would not each such situation be describable as a node, while
the flow of decisions among the possibilities would define a most
extraordinary personal random walking a most complex network?
Although such a dynamic network is undoubtedly out there (or in here),
we are allowed to explore just a small portion of it at a time.  And,
with basis on whatever knowledge we can draw from such a small sample,
we have to decide about critical next steps.  However, the situations
involving incomplete or sampled networks extend much further than this
extreme example.  For instance, the steps along any game or maze is
but a sample of a much larger network of possibilies.  Explorations of
land, sea and space also correspond to small samplings of a universe
of possibilities, not to mention more `classical' large networks such
as those obtained for protein interaction, social contacts and the
Internet.  Last but not least, the own exploratory activities of
science are but a most complex random walk on the intricate and
infinite web of knowledge~\cite{Costa_know:2006}.  In all such cases,
the success of the whole enterprise is critically connected to the
quality and accuracy of the information we can infer about the
properties of the whole network while judging from just a small sample
of it.  Little doubt can be raised about the importance of such a
problem, which has received little attention, except for the
investigations by Serrano et al. on the effects of sampling the WWW
through crawlers~\cite{Serrano_etal:2005}.  The literature about
random walks in complex networks
include~\cite{Tadic:2001,Tadic:2003,Bollt_Avraham:2004,Noh_Rieger:2004}.

The current article is about incomplete and sampled networks and some
related fundamental questions.  We start with the basic mathematical
concepts, identifying some of the most interesting related questions
and perspectives, and proceed by illustrating what can be learnt about
random and Barab\'asi-Albert networks while sampling them locally in
random fashion or through three types of random walks --- traditional
(uniform decision probability) and preferential to untracked edges and
preferential to untracked nodes).

\section{Basic Concepts and Some Fundamental Issues}

An undirected~\footnote{The results in this article are immediately
extended to more general networks, including directed and weighted
models.}  complex network $\Gamma=(\Omega,E)$, involving a set of $N$
nodes $\Omega$ and a set $V$ of $E$ connections between such nodes,
can be completely specified in terms of its $N
\times N$ \emph{adjacency matrix} $K$, such that the existence of a connection
between nodes $i$ and $j$ implies $K(i,j)=K(j,i)=1$ (zero is assigned
otherwise).  The degree $k_i$ of node $i$ is defined as corresponding
to the number of edges connected to that node, and the clustering
coefficient $cc_i$ of that node is calculated as the number of edges
between the neighbors of $i$ and the maximum possible number of
interconnections between these neighbors(e.g.~\cite{Newman:2003}).

\begin{figure}[h]
 \begin{center} 
   \includegraphics[scale=0.3,angle=0]{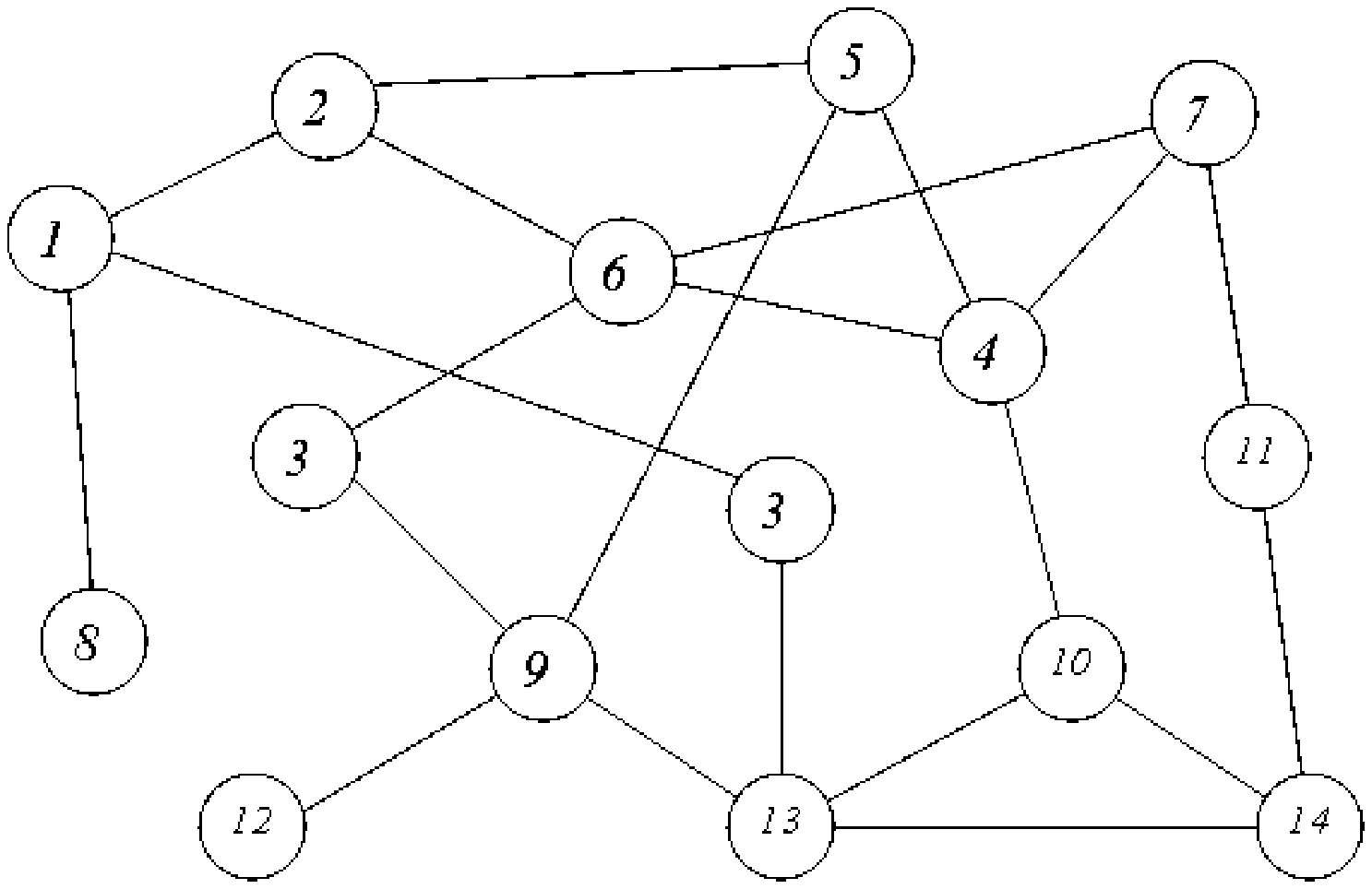} (a) \\
   \includegraphics[scale=0.3,angle=0]{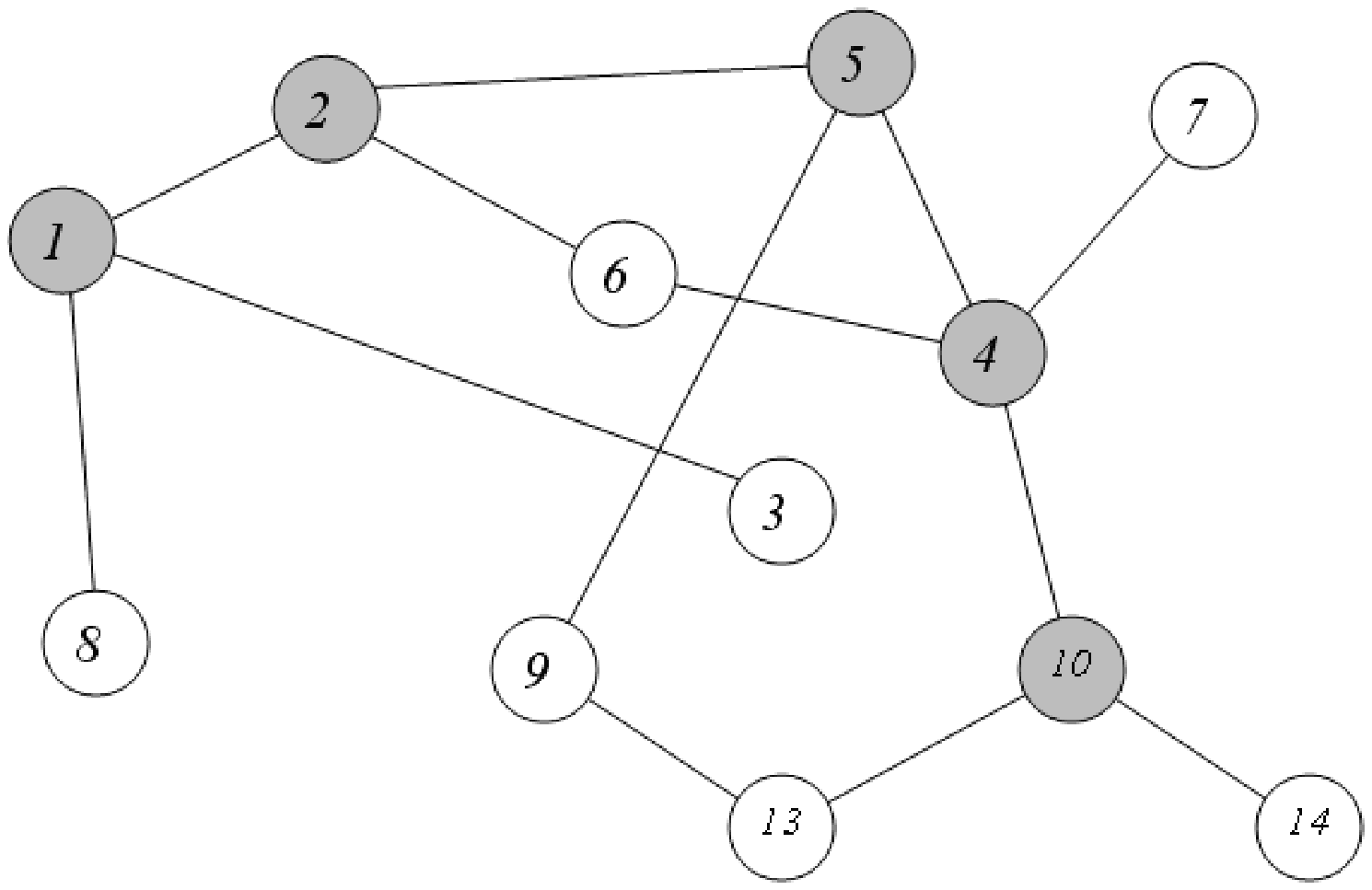} (b) \\
   \includegraphics[scale=0.3,angle=0]{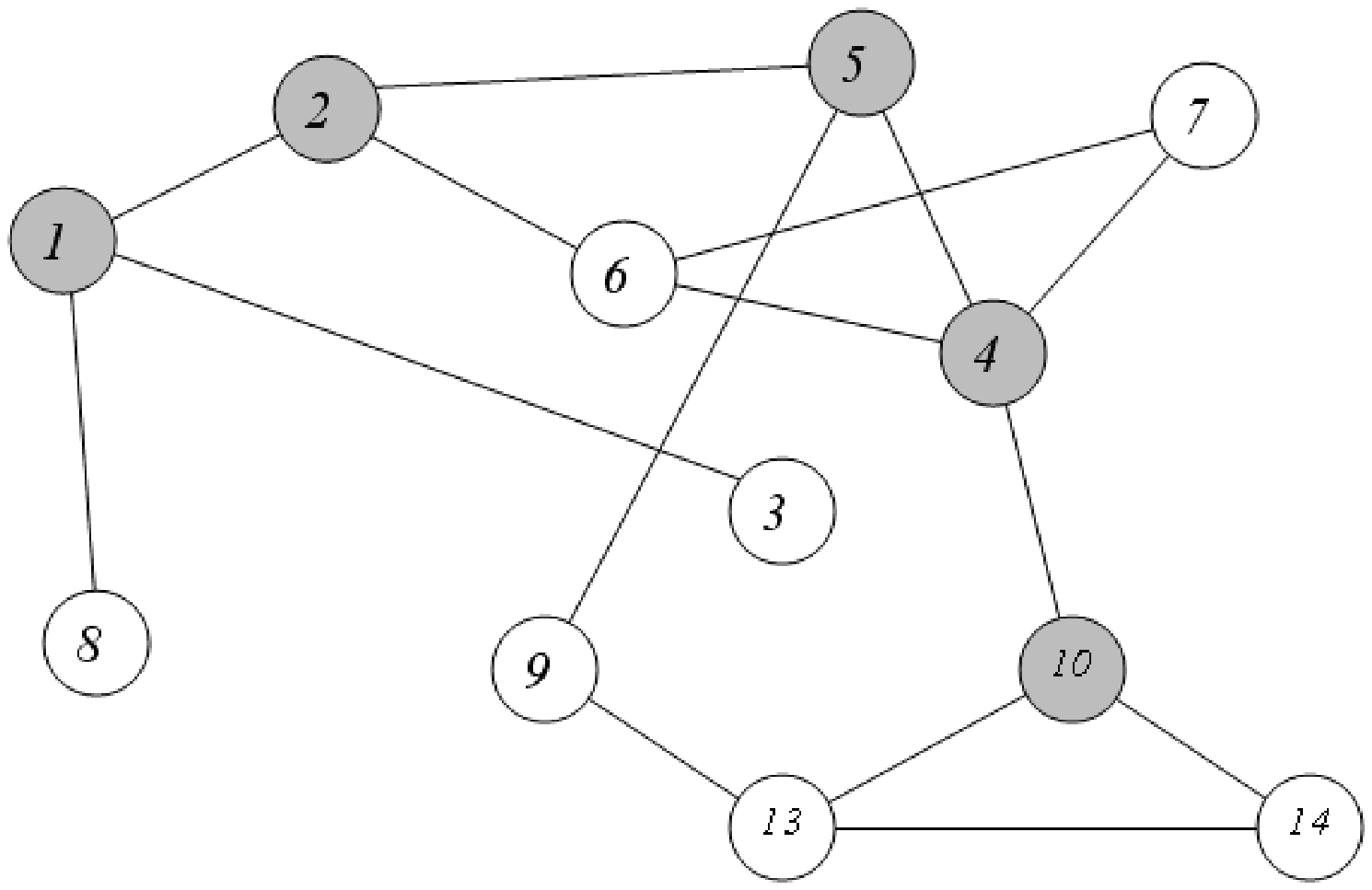} (c) \\
   \vspace{0.5cm} 
   \caption{A simple network (a) and two incompletely specified
   networks obtained by a random walk considering neighboring nodes
   and (b) and the latter plus the edges between adjacent
   nodes (c). The gray nodes correspond to those sampled during the random
   walk~\label{fig:incompl}}
\end{center}
\end{figure}

An \emph{incompletely specified complex network} is henceforth
understood as any subnetwork $G$ of $\Gamma$ such that $G \ne \Gamma$.
In this work we will restrict our attention to incomplete complex
networks defined by sets of nodes and adjacent edges identified during
random walks.  Such networks can be represented as $G = ((i_1, V_1);
(i_2, V_2), \ldots, (i_M, V_M) )$, where $i_p$ are nodes sampled
during the random walk through 
the respective list of adjacent nodes. Note that necessarily $i_{p+1}
\in V_p$ and  that $(i_1, i_2, \ldots, i_M)$ corresponds to a 
\emph{path} along $\Gamma$.  It is also  interesting to consider more 
substantial samples of $\Gamma$, for instance by considering not only
the adjacent edges, but also the interconnections between the
neighboring nodes of each node. Therefore, the case above becomes $G =
((i_1, V_1, E_1); (i_2, V_2, E_2), \ldots, (i_M, V_M, E_M) )$, where
$E_p$ is the set containing the edges between the nodes in $V_p$.
Figure~\ref{fig:incompl} illustrates a complex network (a) and
respective examples of incompletely specified networks obtained by
random walks considering neighboring nodes (b) and the latter plus the
edges between neighboring nodes (c).

Given an incomplete specified complex network $G$, a natural question
which arises is: to what accuracy the properties of the whole network
$\Gamma$ can be inferred from the available sampled information?
Because the estimation of global properties of $\Gamma$ such as
shortest paths and diameter constitutes a bigger challenge to the
moving agent, we concentrate our attention on \emph{local} topological
properties, more specifically the node degree and clustering
coefficient averaged over the network.

Three types of random walks are considered in the present work: (i)
\emph{`traditional':} the next edge is chosen with uniform probability 
among the adjacent edges; (ii) \emph{preferential to untracked edges:}
the next edge is chosen among the untracked adjacent edges and, in
case no such edges exist, uniformly among all the adjacent edges; and
(iii) \emph{preferential to unvisited nodes:} the next edge is chosen
among those adjacent edges leading to unvisited nodes and, in case no
such edges exist, uniformly among all the adjacent edges.  Note that
the plausibility of the preferential schemes depends on each modeled
system.  For instance, the preference to untracked nodes implies that
the moving agent knows whether each edge leads to a still unvisited
node, though it may not know exactly which one.  It is interesting to
note that the process of sampling an existing network through a random
walk can be interpreted as a mechanism for `growing' a network.

\section{Node and Edge Coverage}

First we consider networks growth according to one of the following
two complex network models: (a) \emph{random networks}, where each
pair of nodes has a probability $\lambda$ of being connected; and (b)
\emph{Barab\'asi-Albert networks (BA)}, built by using the preferential 
attachment scheme described in~\cite{Albert_Barab:2002}.  More
specifically, new nodes with $m$ edges each are progressively
incorporated into the growing network, with each of the $m$ edges
being attached to previous nodes with probability proportional to
their respective node degrees. The network starts with $m0=m$ nodes.
Complex networks with number of nodes $N$ equal to $100, 200, \ldots,
900$ and $m=3, 4, \ldots, 8$ have been considered. An equivalent
average degree and number of edges was imposed onto the random
networks as in ~\cite{Costa_know:2006}. A total of 200 realizations of
each configuration, for the three types of random walks, were
performed experimentally.

\begin{figure}[h]
 \begin{center} 
   \includegraphics[scale=0.35,angle=0]{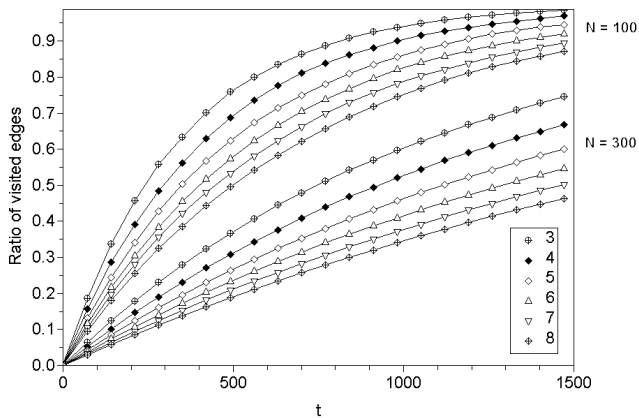} \\
   \caption{The ratio of tracked edges in terms of the steps $t$ for
   $N=100$ and $N=300$ considering the values of
   $m$ as presented in the legend.~\label{fig:curvs}}
\end{center}
\end{figure}

Figure~\ref{fig:curvs} illustrates the ratio of tracked edges in terms
of the steps $t$ for $N=100$ and $N=300$ considering $m=3, 4, \ldots,
8$ and the BA network model.  It is clear from the obtained results
that, as expected, the higher the value of $m$, the smaller the ratio
of visited edges.  Note that the increase of $N$ also contributes to
less efficient coverage of the edges, as expressed by the respective
smaller ratios of visited edges obtained for $N=300$.  For large
enough total number of steps, all curves exhibited an almost linear
initial region followed by saturation near the full ratio of visited
edges (i.e. 1).

\begin{figure}
 \begin{center} 
   \includegraphics[scale=0.5,angle=0]{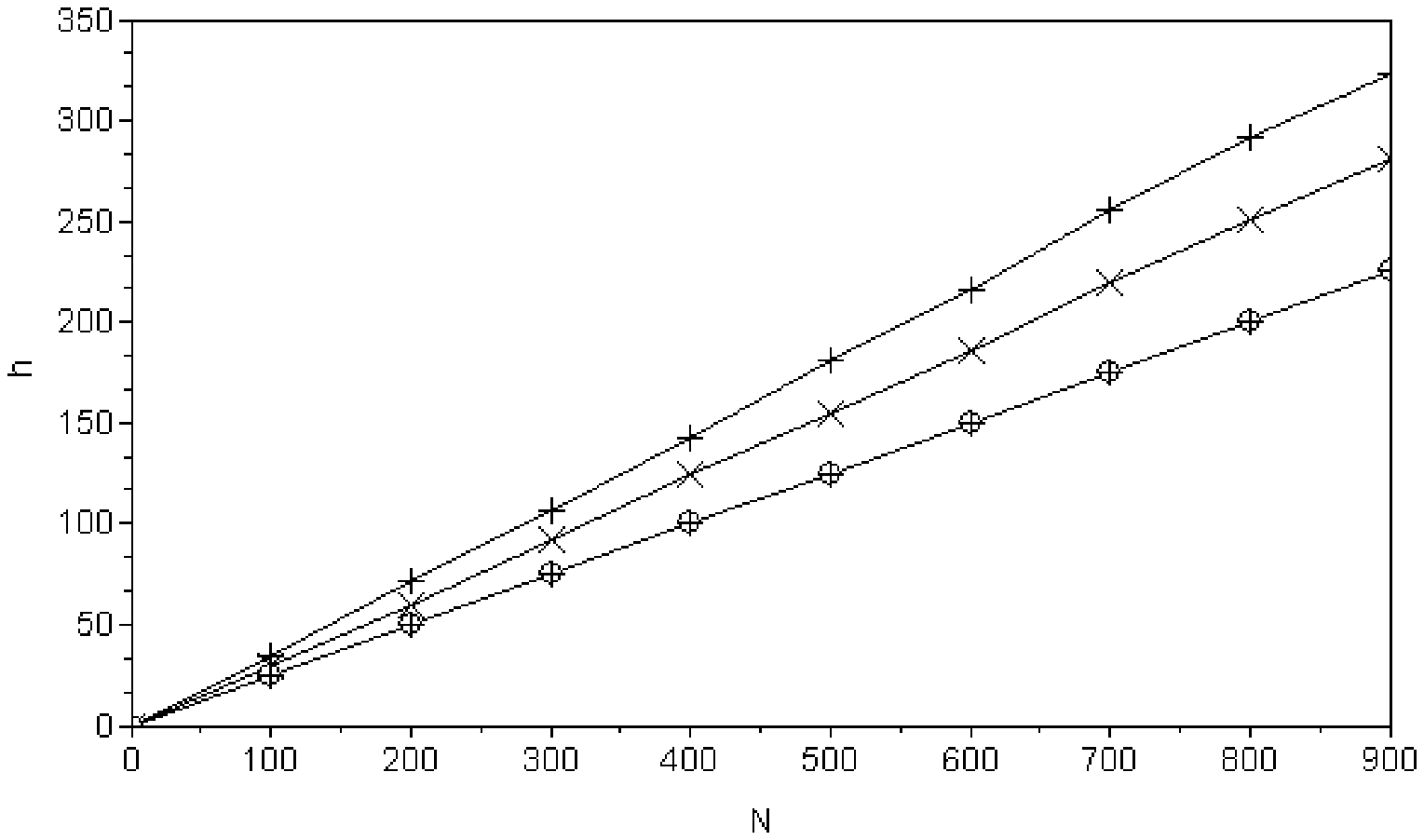} \\
   (a) \\
   \includegraphics[scale=0.5,angle=0]{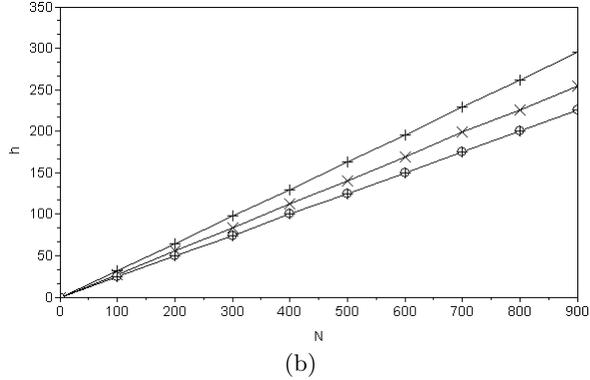} \\
   (b) \\
   \includegraphics[scale=0.5,angle=0]{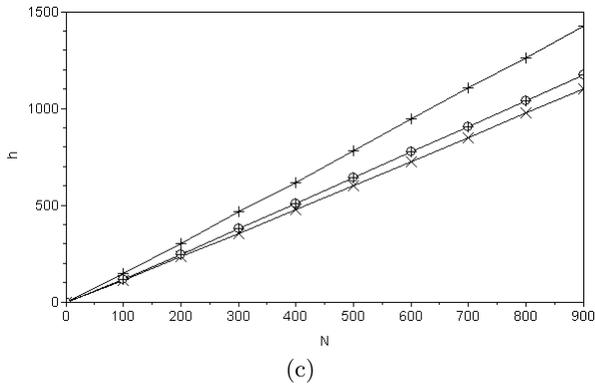} \\
   (c) \\
   \includegraphics[scale=0.5,angle=0]{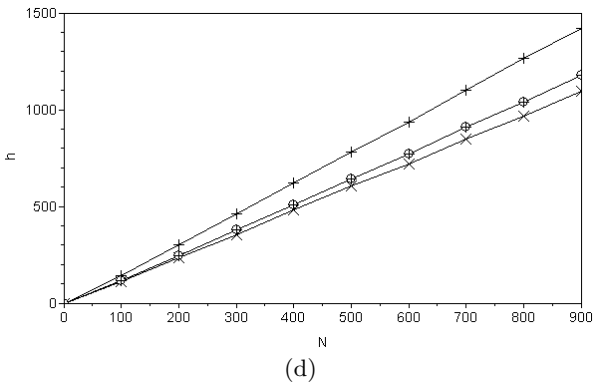} \\
   (d) \\
   \vspace{0.5cm} 
   \caption{The quarter-life of the percentage of visited nodes for BA
   (a) and random (b), and the quarter-life of the percentage of visited
   edges for BA (c) and random (d) models, for traditional (`$+$'),
   preferential to untracked edges (`$\times$') and preferential to
   unvisited nodes (`$\oplus$') random walk
   strategies.~\label{fig:quart_life}}
\end{center}
\end{figure}

Figure~\ref{fig:quart_life}(a) shows the `quarter-lives' $h$ of the
percentage of visited nodes in terms of the network size $N$ with
respect to the BA network model with $m=5$.  This measurement
corresponds to the average number of steps at which the random walk
has covered a quarter of the total number of network nodes.  Similar
results have been obtained for other critical fractions
(e.g. half-life).  Note that, as $m$ is fixed at 5, the average degree
$\left< k \right>$ of all networks in this figure remains equal to
10~\footnote{In the BA model, the average degree is equal to $2m$.},
being therefore constant with $N$, while the average number of edges
grows as $\left< E \right> = N \left< k \right>/2 =
5N$. Interestingly, linear dependence between the quarter-lives and
$N$ are obtained in all cases.  It is also clear from these results
that the most effective coverage of the nodes is obtained by the
random walk preferential to unvisited nodes, with the random walk
preferential to untracked edges presenting the next best performance.
Figure~\ref{fig:quart_life}(b) shows the quarter-lives of the
percentage of visited nodes for random networks.  The random walks
with preference to unvisited nodes again resulted in the most
effective covering strategy.  The quarter-lives for the percentage of
tracked edges are shown in Figures~\ref{fig:quart_life}(c,d)
respectively to BA (c) and random (d) network models.  The best ratios
of covered edges were obtained for the random walk preferential to
untracked edges, with the random walk preferential to unvisited edges
presented the next best performance.  The traditional random walk
resulted the least efficient strategy in all situations considered in
this work.  Note that the three types of random walks resulted
slightly more effective for node coverage in the case of the random
model than in the BA counterparts.

\begin{figure}
 \begin{center} 
   \includegraphics[scale=0.45,angle=0]{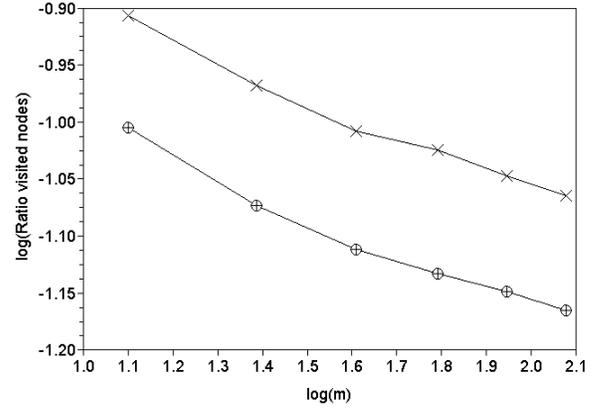} \\ 
   \caption{Loglog
   (Neper) representation of the slopes of the ratios of visited nodes
   obtained for traditional random walks for $m=3, 4, \ldots, 8$
   considering BA (`$\times$') and random (`$\oplus$') network
   models.~\label{fig:slopes}}
\end{center}
\end{figure}

Further characterization of the dynamics of node coverage can be
obtained by considering the scaling of the slopes of the curves of
ratios of visited nodes in terms of several values of $m$.
Remarkably, the slopes obtained by least mean square fitting for the
two types of preferential random walks were verified not to vary with
$m$, being fixed at 0.31 and 0.25, respectively, for the BA model and
0.285 and 0.25 for the random networks. Figure~\ref{fig:slopes} shows
the log-log representation of the slopes in terms of $m$ obtained for
the traditional random walk on BA and random networks for $m= 3, 4,
\ldots, 8$.  It is clear from this figure that, though the slopes tend
to scale in similar (almost linear) fashion for the two types of
considered networks, the ratios of node coverage increase
substantially faster for the random networks.

\section{Estimation of Average Node Degree and Clustering Coefficient}

So far we have investigated the dynamics of node and edge coverage in
random and BA models while considering the three types of random
walks.  In practice, as the size of the network being explored through
the random walks is typically unknown, the number of visited nodes or
tracked edges by themselves provide little information about the
topological properties or nature of the networks.  The remainder of
the present work addresses the estimation of measurements of the local
connectivity of networks, namely the average node degree and average
clustering coefficient, obtalined along the random walks.  

\begin{figure}
 \begin{center} 
   \includegraphics[scale=0.5,angle=0]{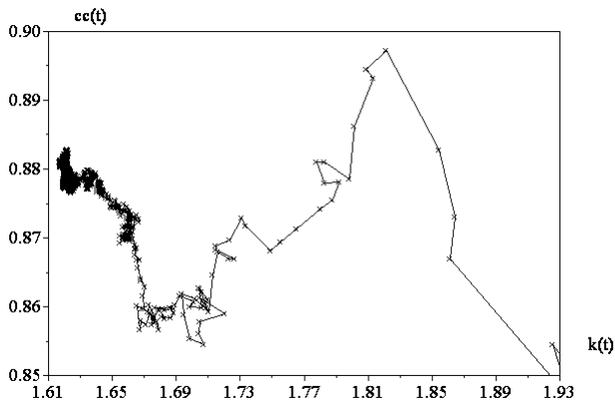} \\ 
   \caption{Curve (actually a kind of random walk) defined by the
   estimations $(k(t),cc(t))$, through traditional random walk, of the
   average node degree $k(t)$ and average clustering coefficient
   $cc(t)$ in a BA network with $N=800$ and $m=5$. The curve
   converges to the incorrect estimations ratios $(1.62,0.88)$ instead
   of $(1,1).$~\label{fig:rw}}
\end{center}
\end{figure}

For generality's sake, the estimations are henceforth presented in
relative terms, i.e. as the ratio between the current estimation
(e.g. $(k(t)$) and the real value (e.g. $\left< k \right>$).
Figure~\ref{fig:rw} illustrates the curve defined by the estimations
$(k(t),cc(t))$ obtained by traditional random walks along a BA network
with $N=800$ and $m=5$.  Interestingly, this curve is indeed a kind of
random walk with convergent limit.  Such curves have been found to
converge to limiting ratios $(k_L,cc_L)$ which can or not correspond
to the ideal ratios $(1,1)$.  In the case of the curve in
Figure~\ref{fig:rw}, we have $(k_L=1.62,cc_L=0.88)$, i.e. the average
node degree has been over-estimated while the average clustering
coefficient has been under-estimated.  Through extensive simulations,
we have observed that the estimations of average node degree tended to
be about twice as much as the real value while the average clustering
coefficient resulted about 0.9 of the real value, irrespectively of
$N$, $m$ or random walk type.  Contrariwise, the estimation of these
two topological features for random networks tended to produce stable
and accurate estimation of the average clustering coefficient, while
the obtained average node degree presented very small variation around
the optimal value.  The substantial biases implied by the random walk
over BA networks is a direct consequence of the larger variability of
node degree exhibited by this model.  Therefore, nodes with higher
degree will tend to be visited more frequently~\footnote{Actually the
rate of visits to the nodes of an undirected complex network, at
thermodynamical equilibrium, can be verified to be proportional to the
node degree.}, implying over-estimation of the average node degree and
a slighted bias on the clustering coefficient.

Provided the moving agent can store all the information obtained from
the network as it is being explored, yielding a partial map of the so
far sampled structure, it is possible to obtain more accurate
(i.e. unbiased) estimates of the average node degree and clustering
coefficient during any of the considered random walks in any type of
networks by performing the measurements without node repetition.
However, an agent moving along a BA network without resource to such
an up to dated partial map will have to rely on averages of the
measurements calculated at each step.  This will cause the impression
of inhabiting a network much more complex (in the sense of having
higher average node degree) than it is indeed the case.  Going back to
the motivation at the beginning of this article, it is difficult to
avoid speculating whether our impression of living in a world with so
many possibilities and complexities could not be in some way related
to the above characterized effects.

\section{Concluding Remarks}

The fact that most real complex networks are only partially available
to us as a consequence of their sheer size and complexity, it becomes
of critical importance to understand how well these structures can be
investigated by using sampling strategies such as different types of
random walks.  The present work has addressed this issue considering
random and BA network models with varying connectivity and sizes being
sampled by three types of random walks.  A series of important results
have been obtained which bear several theoretical and practical
implications.  Particularly surprising is the fact that random and BA
networks are similarly accessible as far as node and edge exploration
is concerned.  Actually, random networks tend to have their nodes and
edges explored in a slightly more effective way.  Also important is
the characterization of linear scaling with the network size of the
quarter-life of the ratio of covered nodes and edges, and the
identification substantial biases in estimations of the average node
degree and clustering coefficient in several situations.  In
particular, the average node degree tend to be estimated as being
approximately twice as large as the real value.  Additional insights
about the non-trivial dynamics of complex network exploration through
random walks can be achieved by considering other network models as
well as more global topological measurements such as shortest paths,
diameters, hierarchical measurements and betweeness centrality.

\vspace{1cm}

Luciano da F. Costa is grateful to CNPq (308231/03-1) for financial
support.

\bibliographystyle{apsrev}
\bibliography{hybr}
\end{document}